\documentclass[reprint,amssymb, amsmath, aps, superscriptaddress, showpacs, footinbib, longbibliography, prb]{revtex4-2}
\usepackage{amssymb}
\usepackage{amsmath}
\usepackage{xcolor}
\usepackage{graphicx,hyperref}
\usepackage{footmisc,booktabs}
\usepackage{gensymb}
\usepackage{sidecap}
\usepackage{microtype}

\begin{document}

\title{
Static and Dynamic Electronic Properties of Weyl Semimetal NbP---A Single Crystal $^{93}$Nb-NMR Study
}

\author{Tetsuro~Kubo}%
\altaffiliation[Present Address: ]{Kanto Yakin Kogyo Co. Ltd., Hiratsuka 254-0014, Japan}
\affiliation{Max Planck Institute for Chemical Physics of Solids, 01187 Dresden, Germany}

\author{Hiroshi~Yasuoka}%
\affiliation{Max Planck Institute for Chemical Physics of Solids, 01187 Dresden, Germany}

\author{Deepa~Kasinathan}%
\altaffiliation[Present address: ]{MHP Management- und IT-Beratung GmbH, Film- und Medienzentrum, K\"onigsallee 49, 71638 Ludwigsburg, Germany}
\affiliation{Max Planck Institute for Chemical Physics of Solids, 01187 Dresden, Germany}

\author{K. M. Ranjith}%
\altaffiliation[Present address: ]{Leibniz Institute for Solid State and Materials Research Dresden, IFW Dresden, Helmholtzstra{\ss}e 20, 01069 Dresden, Germany}
\affiliation{Max Planck Institute for Chemical Physics of Solids, 01187 Dresden, Germany}

\author{Marcus~Schmidt}%
\affiliation{Max Planck Institute for Chemical Physics of Solids, 01187 Dresden, Germany}

\author{Michael~Baenitz}%
\email{michael.baenitz@cpfs.mpg.de}
\affiliation{Max Planck Institute for Chemical Physics of Solids, 01187 Dresden, Germany}

\date{
\today
}%

\begin{abstract}
 Nuclear magnetic resonance (NMR) techniques have been used to study the static and dynamic microscopic properties of the Weyl semimetal NbP. From a complete analysis of the angular dependence of the $^{93}$Nb-NMR spectra in a single crystal, the parameters for the electric quadrupole interactions and the magnetic hyperfine interactions were determined to be $\nu_{\rm Q} = 0.61$\,MHz, $\eta = 0.20$, $(K_{XX}, K_{YY}, K_{ZZ}) = (- 0.06, 0.11, - 0.11)$\% at 4.5\,K. The temperature and field dependence of the $^{93}$Nb Knight shift revealed a characteristic feature of the shape of the density of states with nearly massless fermions. We clearly observed a quantum oscillation of the Knight shift associated with the band structure, whose frequency was in good agreement with the previous bulk measurements. The temperature dependence of the spin-lattice relaxation rate, $1 / T_{1} T$, showed an almost constant behavior for $30 < T < 180$\,K, while a weak temperature dependence was observed below $\sim 30$\,K. This contrasts with the behavior observed in TaP and TaAs, where the $1 / T_{1} T$ measured by the $^{181}$Ta nuclear quadrupole resonance (NQR) shows $1 / T_{1} T \propto T^{2}$ and $T^{4}$ above approximately 30\,K. In TaP, the temperature dependent orbital hyperfine interaction plays a signficant role in nuclear relaxation, whereas this contribution is not observed in TaAs. Two-component spin echo oscillations were observed. The shorter-period oscillation is attributed to the origin of quadrupole coupling, while the longer-period oscillation indicates the presence of indirect nuclear spin-spin coupling, as discussed in other Weyl semimetal like TaP.
\end{abstract}

\maketitle

\section{Introduction}
Since the prediction and experimental realization of topological insulators, topology has played an increasingly significant role in materials science. In topological insulators, the bulk exhibits a gapped insulating state, while unique behaviors emerge in the gapless surface states \cite{Franz2013}.

Recently, a new class of topological materials has gained considerable attention: Dirac, Weyl, and line-nodal semimetals. These materials are distinguished by their topological properties manifesting in the bulk states. Among them, Weyl semimetals have been particularly intriguing. In 2015, Weng \textit{et al.} theoretically predicted the existence of  Weyl fermions in non-centrosymmetric transition metal monopnictides \cite{Weng2015}. This prediction was soon confirmed by the observation of characteristic surface Fermi arcs using angle-resolved photoemission spectroscopy (ARPES) \cite{Lv2015}.

Weyl semimetals have revealed fascinating transport phenomena in their bulk properties. Notably, negative magnetoresistance has been observed, attributed to the chiral anomaly associated with Weyl fermions \cite{Huang2015}. These observations provide evidence for the unique bulk electronic structure in Weyl semimetals. Furthermore, the topology of the Fermi surface has been extensively studied through the angular dependence of Shubnikov--de Haas (SdH) quantum oscillations \cite{Arnold2016}, giving deeper insights into their electronic properties.

NbP is one of the first predicted Weyl monopnictides, demonstrating extremely high positive magnetoresistance and ultrahigh mobility \cite{Shekhar2015}. NbP features two types of  Weyl point pairs, W1 and W2. The W1 pair is situated in the $k_{z} = 0$ plane and is 57\,meV below the Fermi level ($E_{\rm F}$), while the W2 pairs are located in the $k_{z} \approx \pm \pi / 2$ plane and 5\,meV above $E_{\rm F}$ \cite{Klotz2016}. Based on the Fermi surface topology constructed by SdH oscillations, it has been reported that all Weyl points of type W1 are enclosed within one Fermi surface, while all Weyl points of type W2 are enclosed within another Fermi surface. This results in a topologically trivial state \cite{Klotz2016}.

As microscopic probes of bulk electronic states, nuclear magnetic resonance (NMR) and nuclear quadrupole resonance (NQR) techniques offer unique insights. These methods can capture not only the static properties but also the magnetic excitations with the characteristic temperature dependence expected for Weyl fermion excitations. While transport, magnetization, and torque measurements are sensitive to the density of states near the Fermi level and Fermi surface shape, NMR provides additional information through the Knight shift and the nuclear spin-lattice relaxation rate $1 / T_{1} T$. Moreover, the electric field gradient (EFG) offers information about the total real-space distribution of electrons, or equivalently, the occupied states in momentum ($k$) space \cite{Kaufmann1979}.

In this study, we elucidate the static and dynamic magnetic properties of NbP using onsite $^{93}$Nb-NMR spectroscopy. We report on the temperature and field dependence of the Knight shift and $1 / T_{1} T$ together with the spin-echo decay profile, providing new insights into the electronic state of this prototypical Weyl semimetal.

\section{Experimental details}
A single crystal sample of NbP (${\rm size} \sim 4 \times 2 \times 1.5$\,mm$^{3}$, mass = 25\,mg) was prepared by the chemical transport reaction (CTR) method as previously reported elsewhere \cite{Shekhar2015}. The quality of the obtained crystals was checked by electron-probe-microanalysis and powder X-ray diffraction. These data revealed that the crystal is of high quality.

\begin{figure}[ht]
 \centering
 \includegraphics[width=1.00\linewidth]{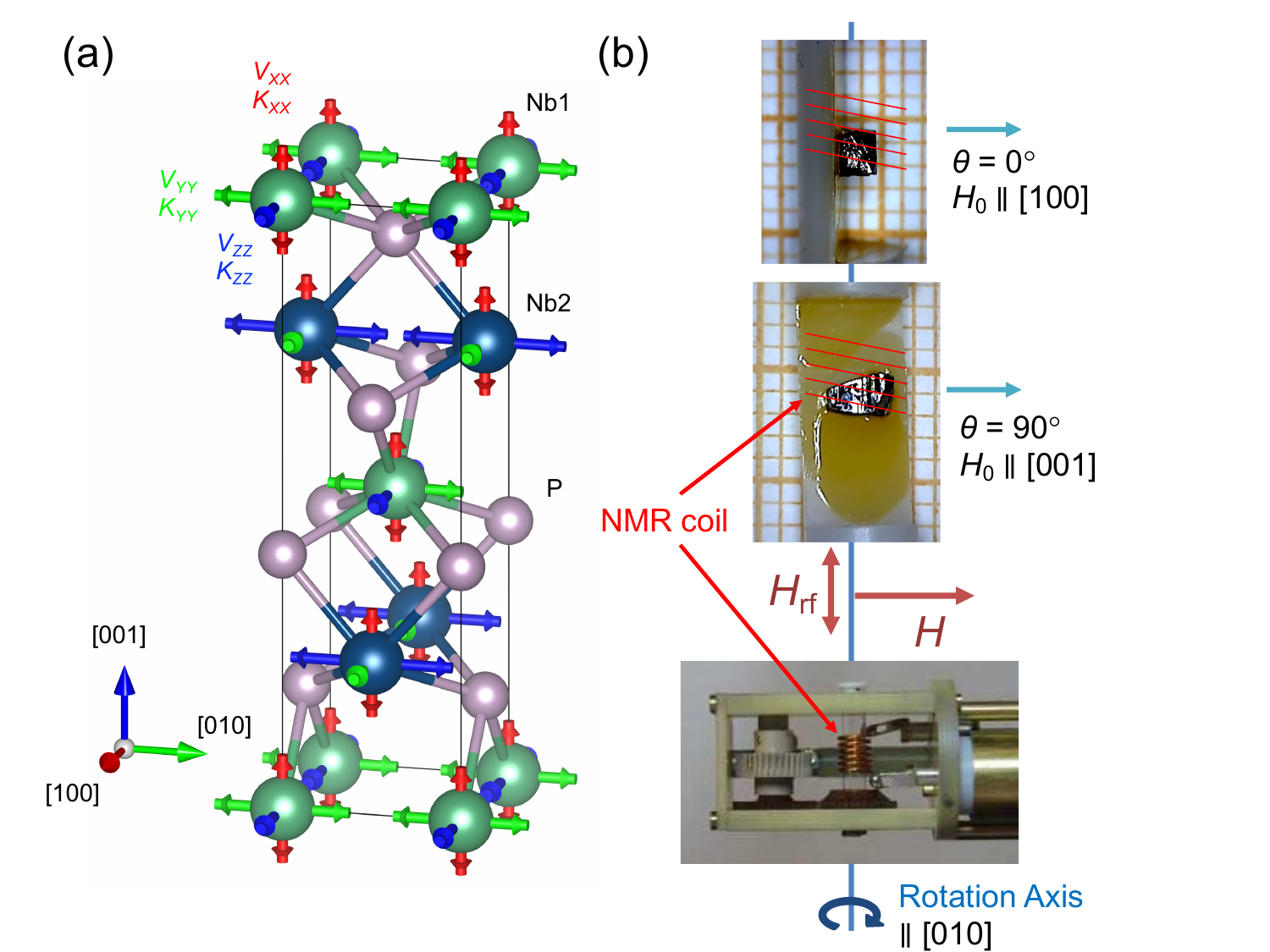}
 \caption{(Color online) Crystal structure and principal axes of the EFG in NbP. Red, green, and blue bars indicate directions of the principal axes of EFG and Knight shift tensors. (b) Setup of the single crystal on the rotator. Here, the crystal was set up with the $[0 1 0]$ axis as the rotation axis. Thus, the external field was rotated from $[0 0 1]$ and $[1 0 0]$. The values in the figure are the goniometer angle $\theta$.}
 \label{fig:structure-setup}
\end{figure}

The crystal structure of NbP belongs to the space group $I 4_{1} m d$ ($C_{4 v}^{11}$, No.\,109) lacking an inversion center as shown in Fig.\,\ref{fig:structure-setup}(a). Regarding the local symmetry of Nb, there is only one Nb site with the maximum principal axes of the Knight shift and EFG tensors along either $[1 0 0]$ or $[0 1 0]$ directions. We define Nb1 as having the $[1 0 0]$ direction as its maximum principal axis and Nb2 as having the $[0 1 0]$ direction as its maximum principal axis. Under the application of the magnetic field, these sites become inequivalent. Note that $[1 0 0]$ and $[0 1 0]$ are equivalent for tetragonal symmetry but should be distinguished when considering the microscopic interactions such as nuclear electric quadrupole or magnetic hyperfine interactions. In order to measure the anisotropy of those interactions, the single crystal was set up with the $[0 1 0]$ axis as the rotation axis as shown in Fig.\,\ref{fig:structure-setup}(b).

The $^{93}$Nb-NMR spectra were measured using a standard pulsed NMR apparatus. As shown in Fig.\,\ref{fig:structure-setup}(b), the sample was mounted with its $[1 0 0]$ axis perpendicular to the stage plane. In this geometry, the goniometer angles $\theta = 0^\circ$ and $90^\circ$ correspond to $H \parallel [1 0 0]$ and $H \parallel [0 0 1]$, respectively. The angular dependence of the NMR spectra was obtained by a field-sweep method, where the field direction was controlled between $\theta = - 22^\circ$ and $+ 108^\circ$ with an angular step of $10^\circ$ using a homemade uniaxial goniometer. To avoid any artificial broadening, the real part of the spin echo was integrated after a proper phase adjustment. The field dependence of the NMR Knight shift was obtained by fast Fourier transform (FFT) of the spin echo at each magnetic field.

To calculate the EFG, band structure calculations were performed using the density functional theory (DFT) solid-state full-potential local-orbital (FPLO) code \cite{Koepernik1999}. Perdew--Wang parametrization of the local density approximation (LDA) for the exchange-correlation functional was used \cite{Perdew1992,Mostofi2008}. The spin-orbit coupling is taken into account by performing full-relativistic calculations, where the Dirac Hamiltonian with a general potential is solved. The quadrupole frequency $\nu_{\rm Q}$ can be obtained from the calculated EFG at the Nb site, which is defined as the second partial derivative of the electrostatic potential $V(\vec{r})$ at the position of the nucleus, $V_{i j} = \partial_{i} \partial_{j} V (0) - \delta_{i j} \Delta V (0)/3$.

The temperature dependence of the NMR relaxation time $T_{1}$ has been measured between 1.5 and 300\,K. The recovery of nuclear magnetization is mainly measured at the $^{93}$Nb central ($I_{z} = + 1/2 \leftrightarrow - 1/2$) transition line for $\vec{H} \parallel [0 0 1]$. Since $T_{1}$ is extremely long below 10\,K ($\sim$several hundred seconds), the progressive saturation method \cite{Mitrovic2001} was used to obtain the nuclear magnetization recovery profile. At higher temperatures, a conventional inversion recovery method was used. At intermediate temperatures, both methods yield the same $T_{1}$ value. The relaxation profile is fitted by the theoretical curve for magnetic relaxations \cite{Narath1967}:
\begin{align}
 M(t)
 &=
 M_{0} \{1 - [1 / 165 \exp (- t / T_{1}) + 24 / 715 \exp (- 6 t / T_{1}) \nonumber \\
 &\quad
 + 6 / 65 \exp (- 15 t / T_{1}) + 1568 / 7293 \exp (- 28 t / T_{1}) \nonumber \\
 &\quad
 + 7938 / 12155 \exp (- 45 t / T_{1})]\}.
 \label{recovery}
\end{align}
Theoretically, it is known that even with finite $\eta$, the relaxation curve follows the above function when the Zeeman interaction is dominant \cite{Chepin1991}. This condition is always satisfied in the present experiments. The obtained data were well fitted by Eq. (\ref{recovery}), indicating that the nuclear relaxation process is dominated by the magnetic fluctuations.

The spin-echo decay profile is obtained by varying the pulse spacing $\tau$ and monitoring the spin-echo amplitude. For the measurement, the repetition time is chosen to be sufficiently longer than $T_{1}$ (typically 8--10 $T_{1}$ at each temperature) to avoid a saturation effect.

\section{Experimental Results and Discussion}
\subsection{NMR spectra and analysis}
\begin{figure}[htbp]
 \centering
 \includegraphics[width=1.00\linewidth]{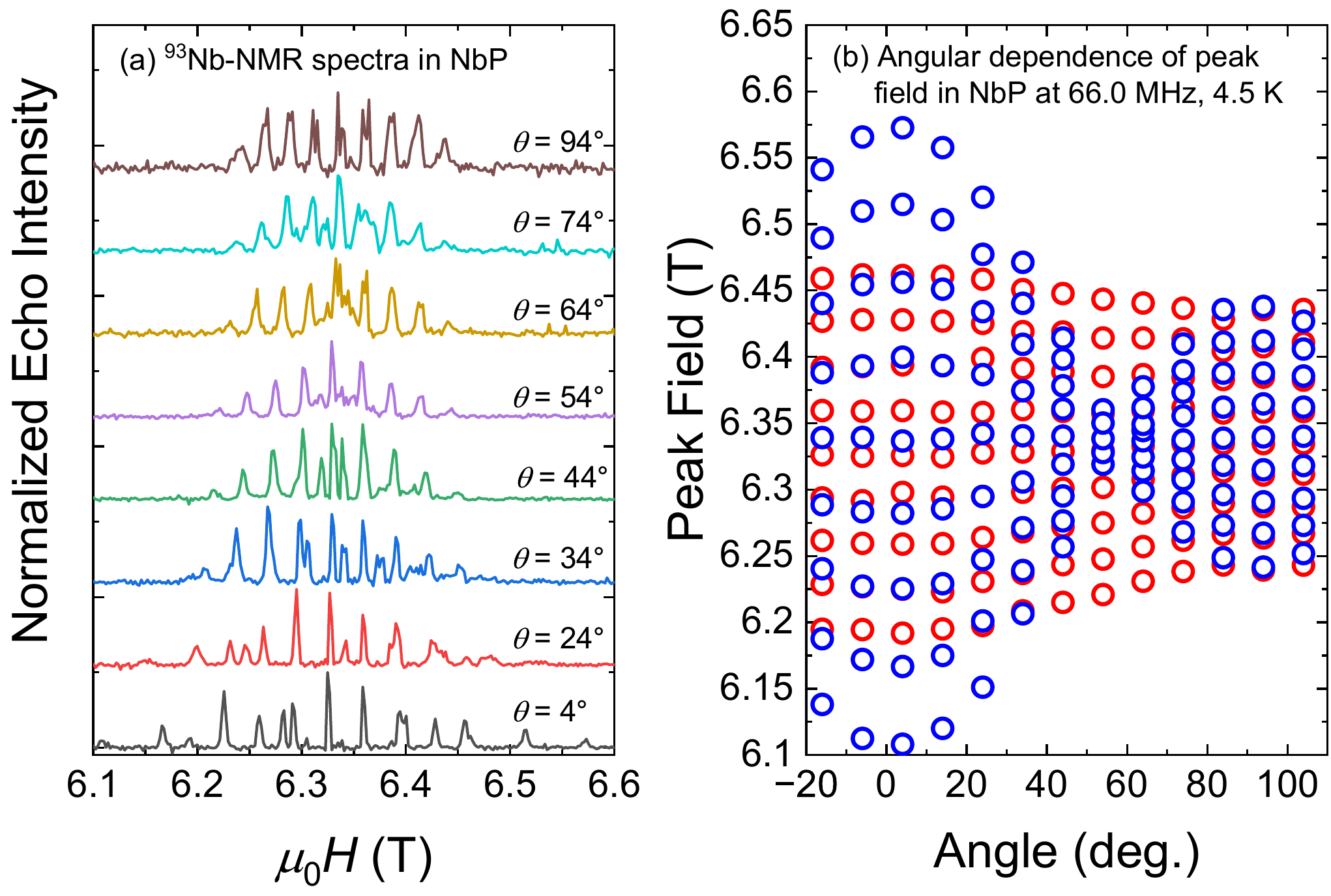}
 \caption{(Color online) Angular dependence of field-swept $^{93}$Nb-NMR spectra (a) and corresponding peak fields (b) in the NbP single crystal observed at 66.0\,MHz and 4.5\,K. The peak fields were obtained from a Gaussian fit of each spectrum shown in Fig.\,\ref{fig:structure-setup}(a).}
 \label{fig:spectrum-peaks}
\end{figure}

\begin{figure}[htbp]
 \centering
 \includegraphics[width=1.00\linewidth]{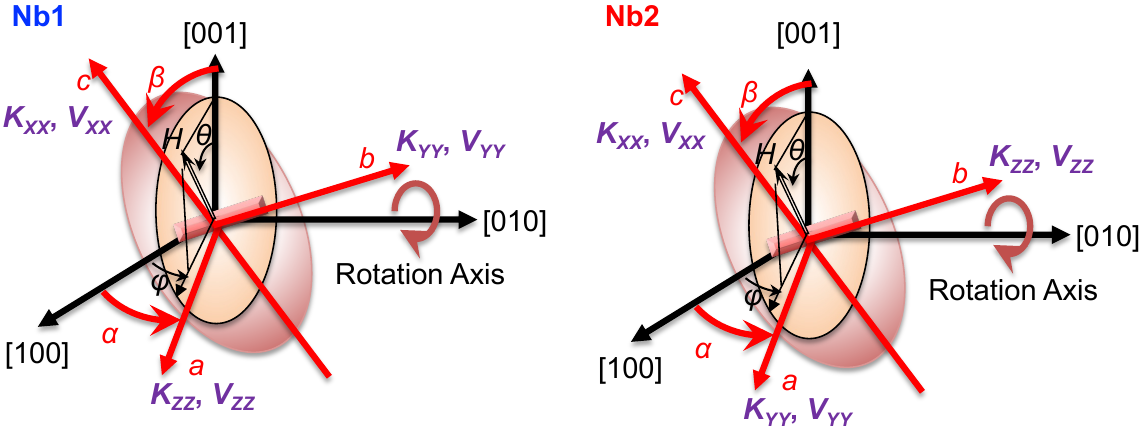}
 \caption{(Color online) Schematic representation of the relationship between the rotation axis and the crystal axes for Nb1 and Nb2, respectively. Here we have introduced angles $\alpha$ and $\beta$ for slight misalignment of the rotation plane. (See Appendix for details.)}
 \label{fig:misalignment}
\end{figure}

We show the angular dependence of the field-swept $^{93}$Nb-NMR spectra in Fig.\,\ref{fig:spectrum-peaks}(a) measured at a fixed frequency of $\nu_{0} = 66.0$\,MHz and 4.5\,K. $\nu_{0}$ corresponds to the resonance field of $\mu_{0} H = 2 \pi \nu_{0} / \gamma_{\rm n} \simeq 6.3$\,T. The spectra were obtained by integrating the real part of the spin-echo signal after an automatic phase control with a field step of 20\,G. The crystal was mounted on the goniometer as shown in Fig.\,\ref{fig:structure-setup}(b). The field direction was chosen to be within the $(0 1 0)$ plane (the rotation axis was $[0 1 0]$). As shown in Figs.\,\ref{fig:structure-setup}(b) and \ref{fig:misalignment}, in this setup the goniometer angles $\theta = 0^{\circ}$ and $90^{\circ}$ ideally correspond to $\mu_{0} H \parallel [1 0 0]$ and $\parallel [0 0 1]$, respectively. A typical linewidth of the central line is narrower than 20\,G (or equivalently $\sim 20$\,kHz), confirming the high quality of the present single crystal.

The quadrupole split spectrum is composed of 18 lines associated with Nb1 and Nb2 sites. The angular dependence can be easily understood as follows. First, NbP has two crystallographically equivalent Nb sites (Nb1 and Nb2) with a local symmetry of $2 m m$ in the unit cell. They are connected by a $4_{1}$-screw rotation. Because of this local symmetry, they share one of the three principal axes of the EFG and Knight shift tensors, $\hat{V}$ and $\hat{K}$, parallel to the $[0 0 1]$ direction. The left two axes rotate by $90^{\circ}$ in the $(0 0 1)$ plane. Then, we expect nine NMR lines for $B_{0} \parallel [0 0 1]$ and 18 (nine lines $\times$ two sites) for $\mu_{0} H \parallel [1 0 0]$. Note that under this rotation condition, the quadrupole satellite lines for Nb2 should exhibit weak angular dependence for small $\eta$, as was observed experimentally. In Fig.\,2(b), we show the angular dependence of the NMR peak fields obtained by Gaussian fitting of all the lines obtained. As can be seen in the figure, we can divide the peaks into two groups, one with stronger field-angle dependence (denoted by blue circles) and another with weaker field-angle dependence (red circles), each consisting of nine peaks.

\begin{figure}[htbp]
 \centering
 \includegraphics[width=1.00\linewidth]{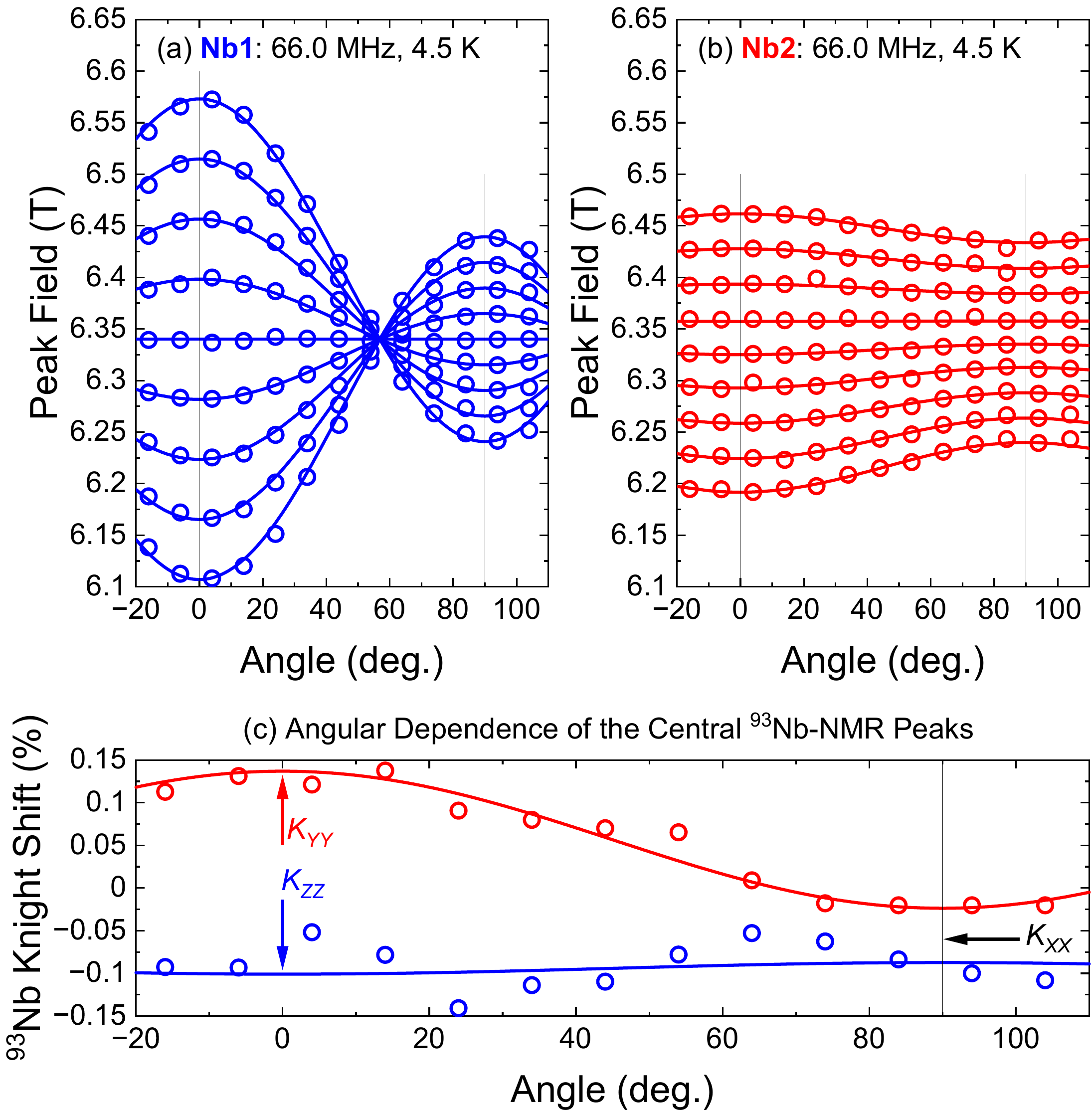}
 \caption{(Color online) Extracted angular dependence of quadrupolar spilt satellite lines for Nb1 (a, blue) and for Nb2 (b, red). (c) shows the angular dependence of the Knight shift obtained from the central transition, from which the anisotropy parameter of the Knight shift can be extracted.}
 \label{fig:peaks}
\end{figure}

To make the angular dependence clearer, the expanded plots for the quadrupole satellites and the central transition of Nb1 and Nb2 are shown in Fig.\,\ref{fig:peaks}(a), (b), and (c).

As clearly shown in Fig.\,\ref{fig:peaks}(c), two peaks are observed at $\theta = 90^{\circ}$ ($\mu_{0} H \parallel [0 0 1]$). This observation suggests that the rotation plane should be slightly different from the $(0 1 0)$ plane.

In general, the resonance field $H_{\rm res}$ for $I_{z} = m$ and $m - 1$ with field angles $\theta$ and $\phi$ can be expressed as,
\begin{align}
 H_{\rm res}
 & =
 \frac{H_{\rm L}}{1 + K_{\rm iso} + K_{\rm ax} (3 \cos^{2} \theta - 1 + \epsilon \sin^{2} \theta \cos 2 \phi)} \nonumber \\
 & +
 \frac{\pi \nu_{\rm Q}}{\gamma_{\rm n}} (3 \cos^{2} \theta - 1 + \eta \sin^{2} \theta \cos 2 \phi) (m - 1 / 2), \\
 H_{\rm{L}}
 & =
 \frac{2 \pi \nu_{0}}{^{93}\gamma_{\rm n}}, \quad \nu_{\rm Q} = \frac{e Q V_{ZZ}}{24 h}\\
 K_{\rm iso}
 & =
 \frac{K_{XX} + K_{YY} + K_{ZZ}}{3}, \\
 K_{\rm ax}
 & =
 \frac{2 K_{ZZ} - K_{XX} - K_{YY}}{3},
\end{align}
for small Knight shift and quadrupole interactions, where the first term comes from the Zeeman interaction and the second from the electric quadrupole interaction \cite{Cohen1957,Carter}. Both terms can be expressed including an asymmetry parameter, $\epsilon$ or $\eta$, due to the magnetic and quadrupole interactions as,
\begin{align}
 \epsilon = \frac{K_{XX} - K_{YY}}{K_{ZZ}}, \\
 \eta = \frac{V_{XX} - V_{YY}}{V_{ZZ}},
\end{align}
where $^{93}\gamma / 2 \pi = 10.42107$\,MHz/T is the nuclear gyromagnetic ratio of $^{93}$Nb nuclei and $\nu_{\rm Q}$ is the quadrupole frequency between the nuclear quadrupole moment ($^{93}Q = - 0.320 \times 10 ^ {- 28}$ m$^ {2}$) and the EFG at the Nb site.

The blue and red solid lines in Fig.\,\ref{fig:peaks}(a--c) are the calculated angular dependences of the peak fields by numerical diagonalization of the nuclear spin Hamiltonian. Details of the analysis are given in the Appendix. From the angular dependence of the peak fields, we can extract the NQR parameters as $\nu_{\rm Q} = 0.61 \pm 0.01$\,MHz and $\eta = 0.20 \pm 0.02$, with $V_{XX} \parallel [0 0 1]$, $V_{Y Y} \parallel [0 1 0]$ ($\parallel [1 0 0]$), and $V_{Z Z} \parallel [1 0 0]$ ($\parallel [0 1 0]$) for the Nb1 (Nb2) site. The experimental values agree within $\sim 10$\% with the DFT calculation for NbP, as shown in Table \ref{tab:NQRparam} together with the previous result for TaP \cite{Yasuoka2017}, confirming the validity of the determined NQR parameters and principal axis directions.

\begin{table}[h]
 \caption{\label{tab:NQRparam} Summary of NMR and NQR parameters in NbP and TaP \cite{Yasuoka2017}. Experimental values are compared to the DFT calculation. Note: the maximum principal axis of the EFG is parallel to $[1 0 0]$ for the Nb1 (Ta1) site and $[0 1 0]$ for the Nb2 (Ta2) site in the unit cell.
}
 \begin{tabular}{ccccccc} \hline
  & & $V_{X X}$  & $V_{Y Y}$ & $V_{Z Z}$ & $\nu_{\rm Q}$ & $\eta$ \\
  & & & ($10^{21}$ V/m$^{2}$) & & (MHz) & \\ \hline
  NbP & Cal. & $- 0.837$ & $- 1.233$ & $2.070$ & $0.667$ & $0.1913$ \\
  & Exp. & & & & $0.61$ & $0.20$ \\
  TaP & Cal. & $- 1.186$ & $- 2.354$ & $3.540$ & $19.380$ & $0.3299$ \\
  & Exp. & & & & $19.250$ & $0.423$ \\ \hline
 \end{tabular}
\end{table}

We also determined the principal values of $\hat{K}$ as $(K_{X X}, K_{Y Y}, K_{X X} = (- 0.06 \pm 0.03, 0.11 \pm 0.02, - 0.11 \pm 0.02)$\% from the field shift of the central peaks as shown in Fig.\,\ref{fig:peaks}(c). In addition, we find a misalignment angle, $\alpha$, of about 6 degrees.

\subsection{Temperature and field dependence of the Knight shift}
\begin{figure}[htbp]
 \centering
 \includegraphics[width=1.00\linewidth]{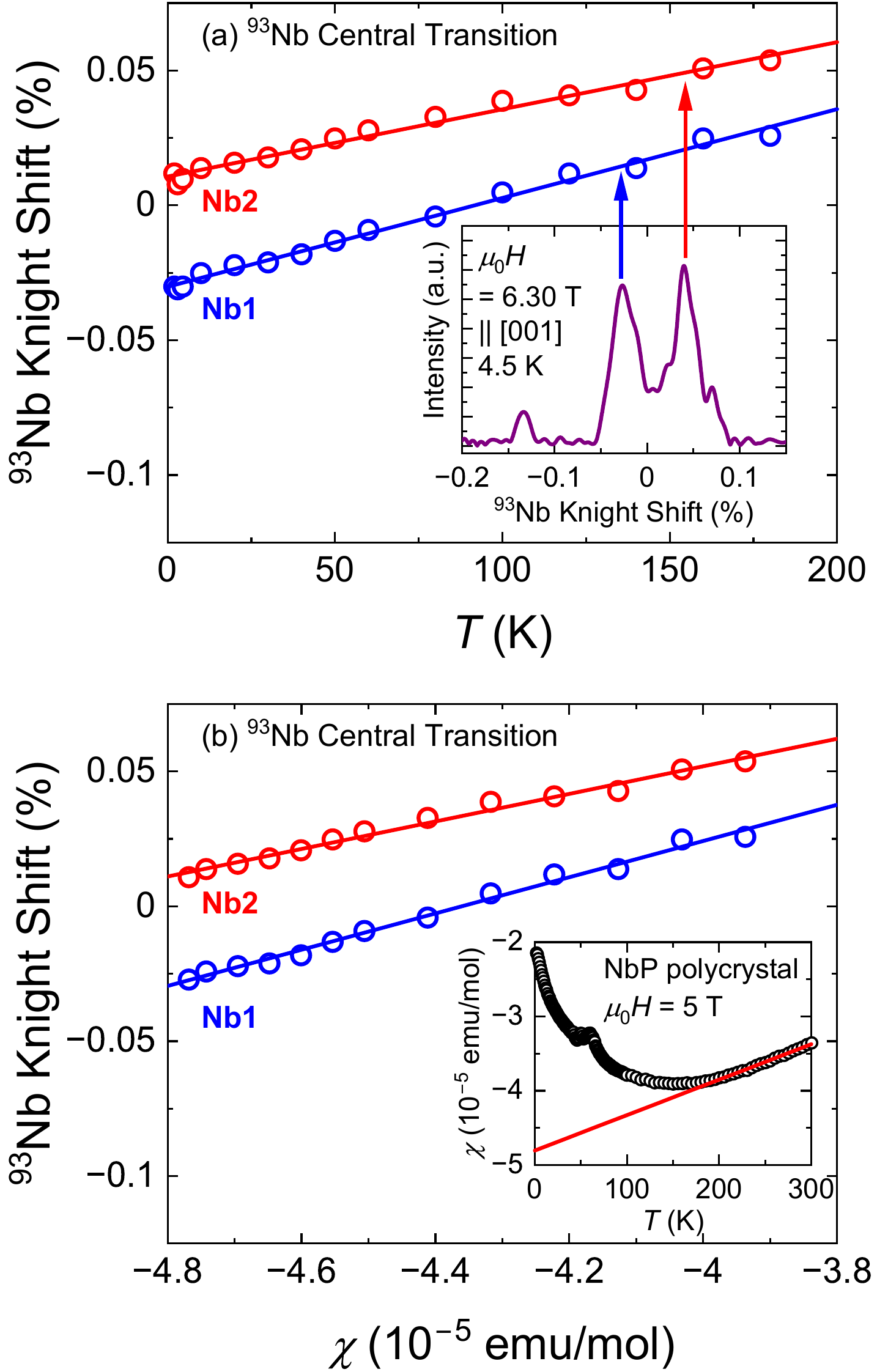}
 \caption{(Color online) Temperature dependence of the $^{93}$Nb Knight shift for Nb1 (red) and Nb2 (blue) for $\mu_{0} H \parallel [0 0 1]$ axis. The inset shows a typical spectrum at 4.5\,K and the two Nb sites are indicated. (b) The Nb Knight shift vs powder susceptibility plot with temperature as an implicit parameter. The slopes give us hyperfine constants of 370 and 280 kOe/$\mu_{\rm B}$ for Nb1 and Nb2, respectively. The inset shows the susceptibility of a polycrystal as a function of temperature. The kink at 55\,K is due to oxygen.}
 \label{fig:shift-K-chi}
\end{figure}

In Fig.\,\ref{fig:shift-K-chi}(a), we show the temperature dependence of the Knight shift measured at the $^{93}$Nb central transition with $\mu_{0} H$ almost parallel to the $[0 0 1]$ direction. In this direction, we expect only one NMR site and measure $K_{XX}$. Experimentally, however, we always observed two central lines in the field and temperature range due to a slight misalignment of the sample.

As shown in Fig.\,\ref{fig:shift-K-chi}(a), within the measured temperature range (4--180\,K), $K$ shows a linear temperature dependence, suggesting the Landau quantized two-dimensional density of states, $D (E) \sim E$, as was observed by the $^{75}$As-NMR Knight shift in TaAs \cite{Wang2020}.

Although measuring the intrinsic magnetic susceptibility directly is challenging, we can estimate it by extrapolating the $\chi$ data above 200\,K, as shown by the red solid line in the inset of Fig.\,\ref{fig:shift-K-chi}(b). This allows us to plot $K$ versus $\chi$ with temperature as the implicit parameter, as depicted in Fig.\,\ref{fig:shift-K-chi}(b). The slope of the $K$ versus $\chi$ plot yields the hyperfine coupling constants, $A_{\rm hf} = 370$ and 280\,kOe/$\mu_{\rm B}$ for Nb1 and Nb2, respectively. Generally, the Knight shift reflects the intrinsic ``local susceptibility'' in concern. While the bulk susceptibility estimated by this linear fitting is provisional and is obtained for a polycrystalline sample, the magnitude of the obtained coupling constants provides a resonable estimate.

\begin{figure}[htbp]
 \centering
 \includegraphics[width=1.00\linewidth]{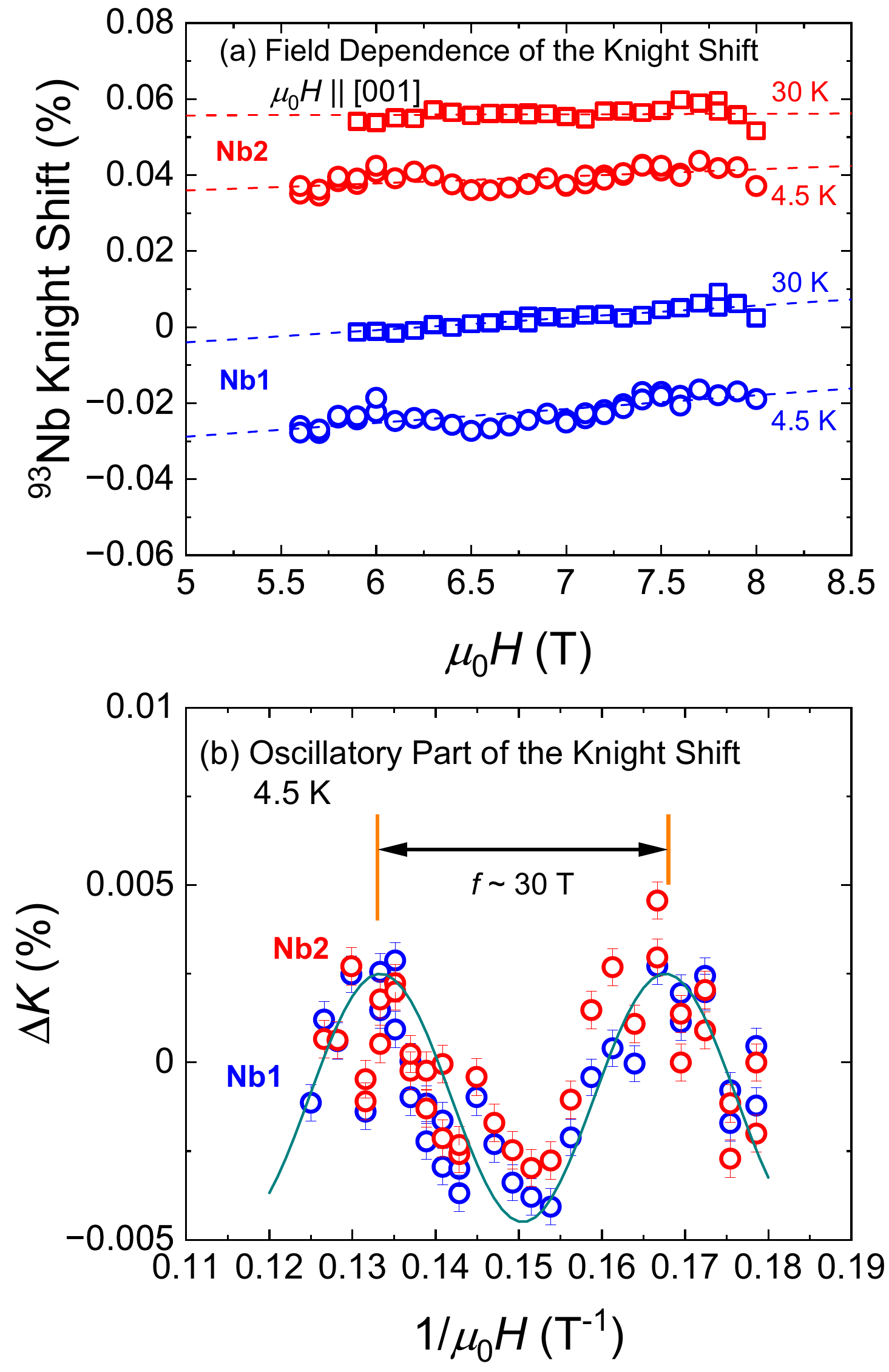}
 \caption{(Color online) (a) Field dependence of the oscillatory behavior of the $^{93}$Nb Knight shift. (b) The oscillatory part indicates a frequency for the quantum oscillation corresponding to 30\,T as in the bulk measurement.}
 \label{fig:shift-qo}
\end{figure}

Note that obtaining large values does not necessarily solely reflect the spin part; it may also include an unknown orbital hyperfine contribution. To understand microscopically the nature of the Weyl fermions in the magnetic field, the field dependence of the Knight shift was measured between 5.5 and 8.0\,T and at 4.5 and 30\,K for $\mu_{0} H \parallel [0 0 1]$. The results are shown in Fig.\,\ref{fig:shift-qo}. As can be seen in the figure, the Knight shift clearly shows oscillatory behavior. Such oscillation is observed in simple metals such as Sn \cite{Khan1967}, Cd \cite{Khan1970}, and Al \cite{Goodrich1971}. They pointed out that, unlike the usual quantum oscillation (QO) measurements, the Knight-shift oscillation contains information about the electronic wavefunctions that couple the electrons and the NMR nuclei.

To gain further insight into the oscillatory behavior observed in $K (\mu_{0} H)$, we subtracted the $H$-linear background and plotted the difference $\Delta K$ against $1/\mu_{0} H$. The result is shown in Fig.\,\ref{fig:shift-qo}(b). We can see the clear oscillations with a frequency of $\sim 30$\,T. This frequency agrees within $\sim 10$\% with the $\alpha_{1}$ orbit around the E1 electron pocket which includes a pair of Weyl points W1 $\sim 57$\,meV below as determined by the de Haas-van Alphen (dHvA) measurement \cite{Klotz2016}. This is one of the first NMR observation of QO in topological semimetals, suggesting the usefulness of NMR to observe the Landau quantized electronic states.

\subsection{Spin-lattice relaxation rate $1 / T_{1} T$}
$1 / T_{1} T$ is related to the wave-vector ($q$) and energy ($\omega$) dependent dynamical spin susceptibility $\chi (q, \omega)$ and is expressed as \cite{Abragam,Slichter},
\begin{equation}
 \frac{1}{T_{1} T} = \frac{2 \gamma_{\rm n}^{2} k_{\rm B}}{g^{2} \mu_{\rm B}^{2}} \sum_{q} A_{q} A_{- q} \frac{{\rm Im} \chi_{\perp} (q, \omega_{0})}{\omega_{0}}
\end{equation}
where $A_{q}$ is the $q$-dependent hyperfine coupling constant and ${\rm Im} \chi_{\perp} (q, \omega_{0})$ is the imaginary part of the transverse component of $\chi (q, \omega)$. Therefore, if we know the $A_{q}$, the temperature dependence of $1 / T_{1} T$ provides us with low frequency magnetic excitations, since the resonance frequency, $\omega_{0}$, is typically several MHz (several $10^{-5}$\,meV). The experimental results measured at $\mu_{0} H = 6.6$ and 3.6\,T are shown in Fig.\,\ref{fig:shift-qo}. It can be seen that $1 / T_{1} T$ above about 10\,K reveals the Korringa process ($1 / T_{1} T = {\rm const.}$) for the low-lying excitations.

\begin{figure}[htbp]
 \centering
 \includegraphics[width=1.00\linewidth]{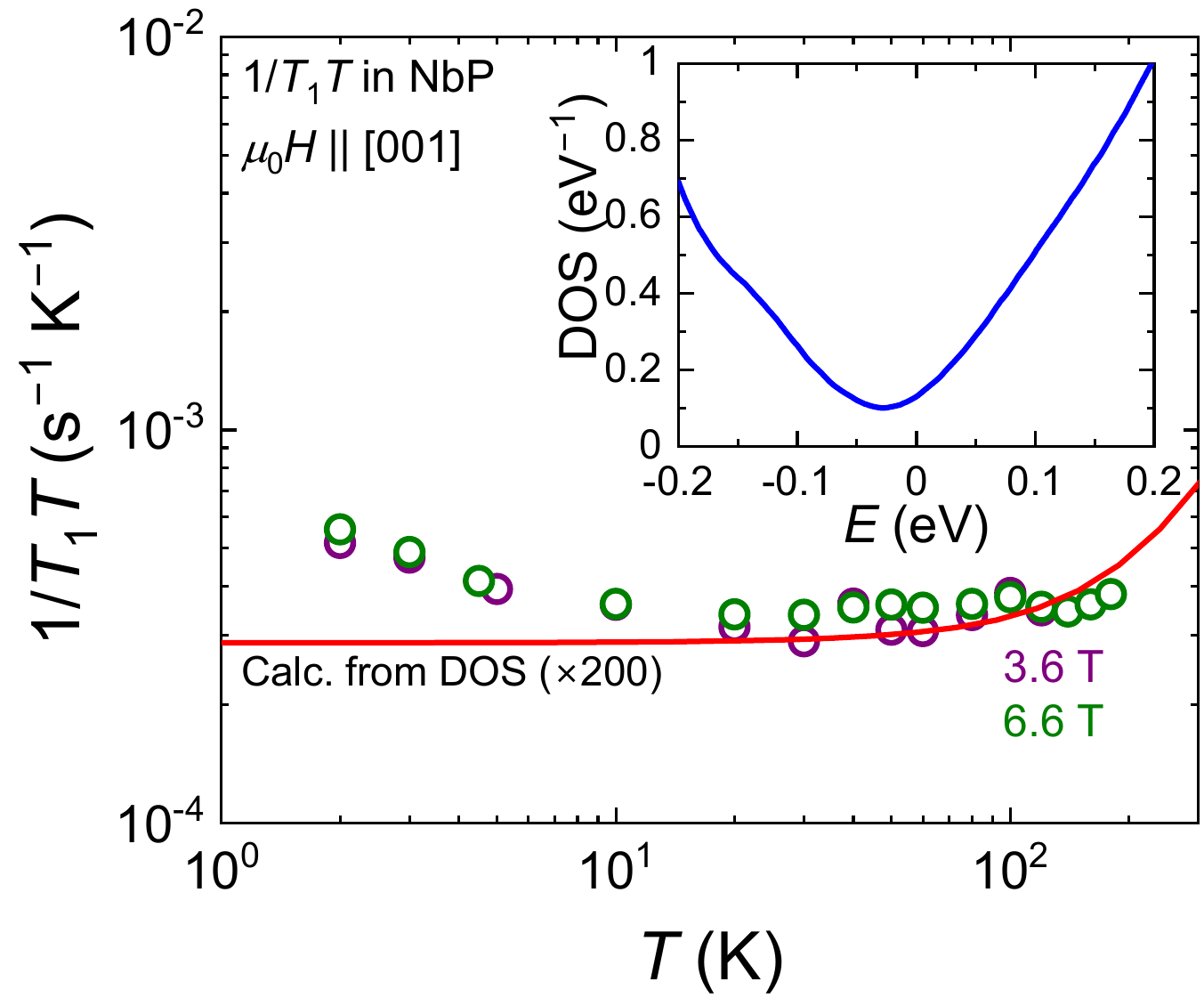}
 \caption{(Color online) Temperature dependence of the $^{93}$Nb-NMR spin-lattice relaxation rate, $1 / T_{1} T$, at 6.5 and 3.6\,T. There is no significant difference between the data at 6.6 and 3.3\,T. The solid red curve is a calculated $1 / T_{1} T$ from DOS shown in the inset. The calculated values are normalized to the experimental values around 70\,K (see text for more details).}
 \label{fig:T1-DOS}
\end{figure}

To understand the high-$T$ Korringa behavior, we calculated $1 / T_{1} T$ based on the density of states in non-interacting systems. Since we do not know the reliable knowledge for the spin part of the hyperfine coupling constant, $A_{\rm hf}$, we set it to be one. Then, $1 / T_{1} T$ is given by \cite{Abragam,Slichter},
\begin{equation}
 \frac{1}{T_{1} T} = \frac{1}{T} \int D^{2} (E) f (E) [1 - f (E)] {\rm d} E,
\end{equation}
where $D (E)$ is the density of states, and $f (E)$ is the Fermi-Dirac function. The calculated result is shown by the red curve in Fig.\,\ref{fig:T1-DOS} with the DOS of NbP in the inset of Fig.\,\ref{fig:T1-DOS}. The temperature dependence of $1 / T_{1} T$ is qualitatively reproduced by the non-interacting DOS model, suggesting that the Weyl fermion excitations in $1 / T_{1}$ were not observed, at least below the highest temperature we measured, 180\,K.

\begin{figure}[htbp]
 \centering
 \includegraphics[width=1.00\linewidth]{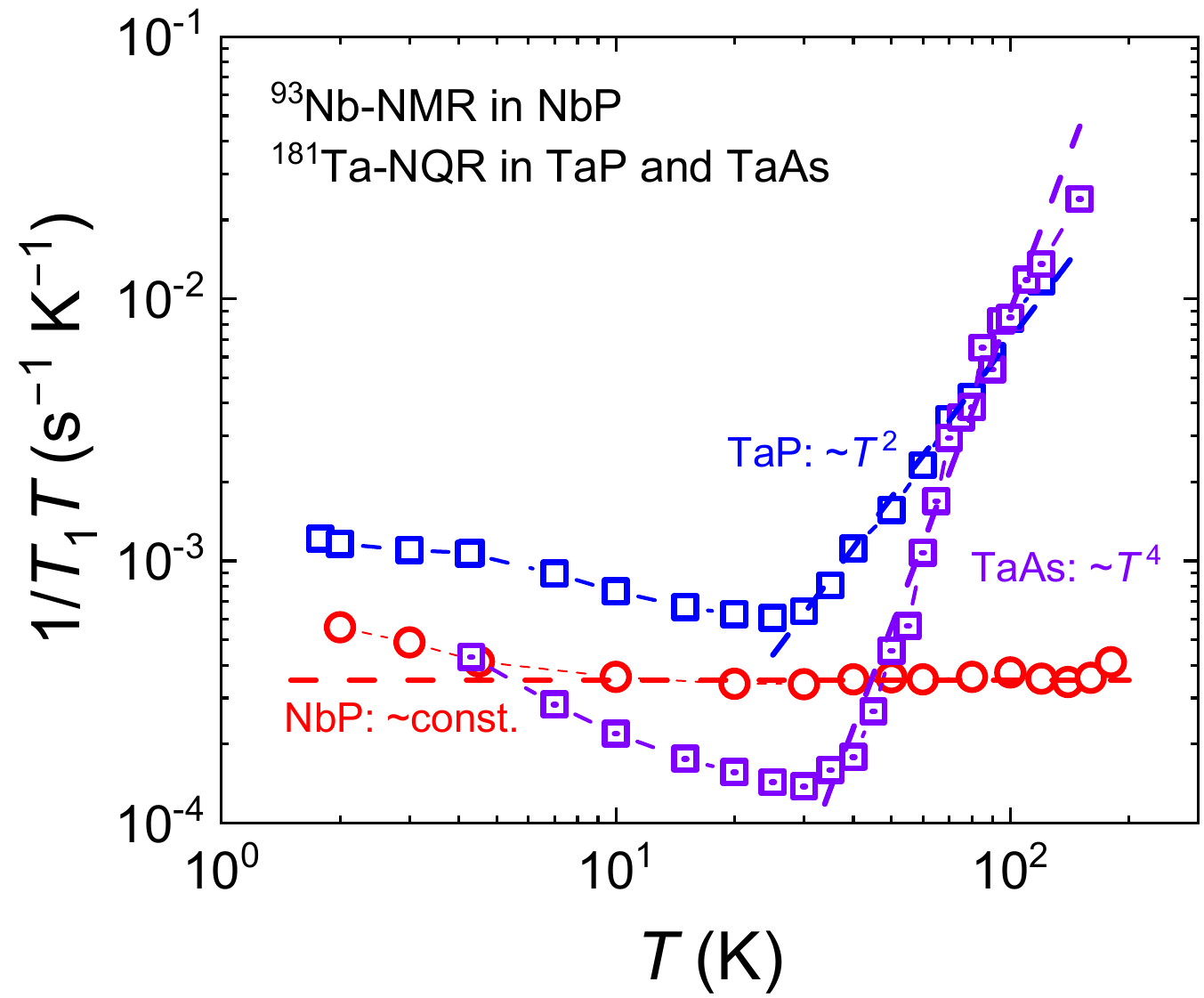}
 \caption{(Color online) Comparison of the temperature dependence of $1 / T_{1} T$ in typical Weyl semimetals, including TaP, TaAs, and NbP, shows distinct characteristics. In TaP and TaAs, $^{181}$Ta-NQR measurements were made at zero magnetic field. In contrast, for NbP, $1 / T_{1} T$ was obtained by $^{93}$-NMR measurements under an external magnetic field.}
 \label{fig:T1-comp}
\end{figure}

In previous NMR/NQR studies of Weyl semimetals, such as TaP and TaAs, $1 / T_{1} T$ showed an abrupt upturn at 20--30\,K and was followed by a power law, $T^{\beta}$, characteristic of excitations in the point-nodal Weyl fermion band. While $\beta = 4$ is naively expected for point-nodal case as TaAs \cite{Kubo2023}, the observed $\beta = 2$ behavior for TaP was attributed to a unique temperature dependence of the orbital hyperfine coupling \cite{Yasuoka2017}. These contrasting power-law behaviors are illustrated in Fig.\,\ref{fig:T1-comp} along with the case for NbP.

\subsection{Spin-echo oscillation}
\begin{figure}[htbp]
 \centering
 \includegraphics[width=1.00\linewidth]{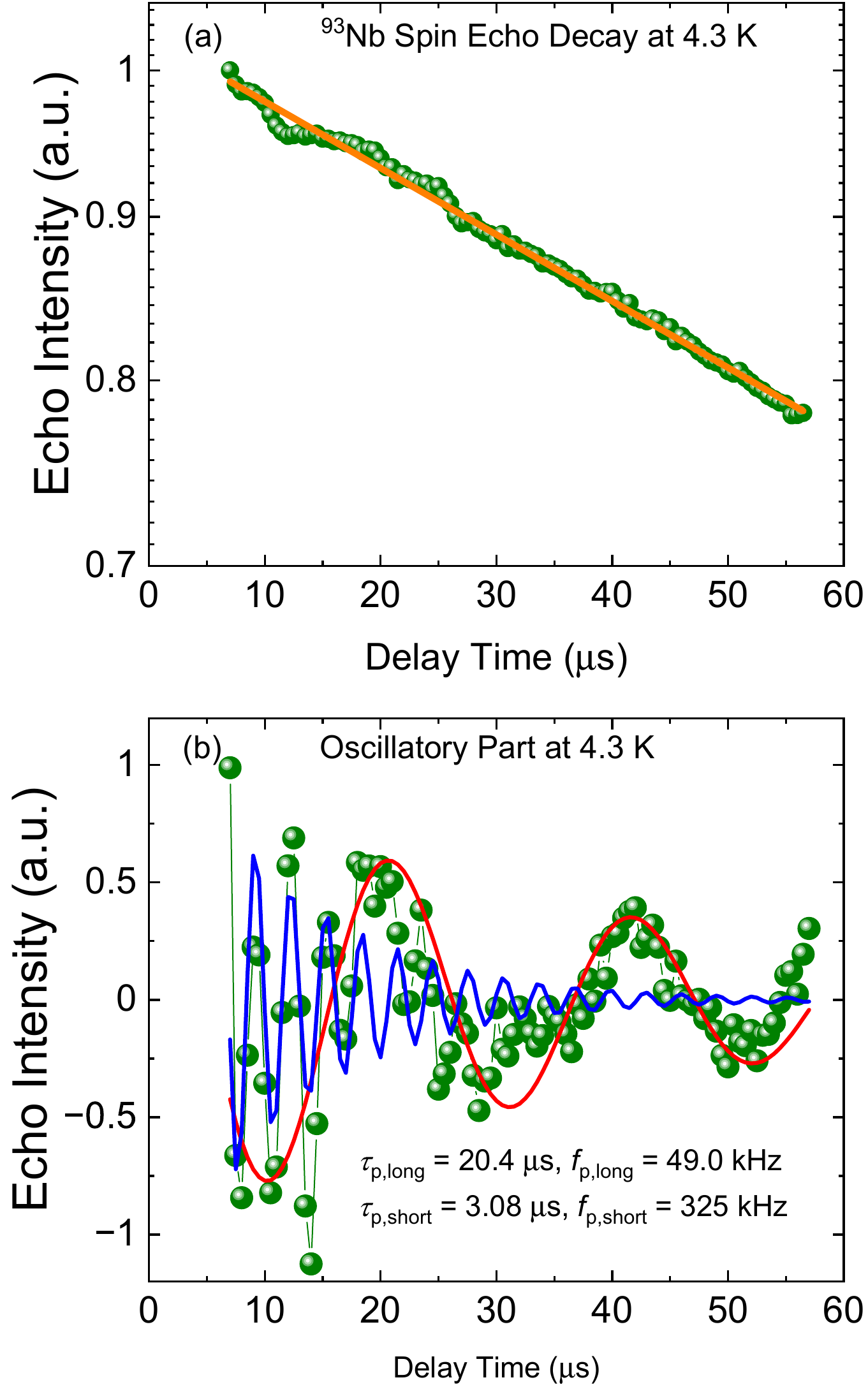}
 \caption{(Color online) $^{93}$Nb-NMR spin-echo decay as a function of the delay time between first and second pulses at 4.3\,K. (b) Oscillatory part of the spin-echo decay. Two oscillation frequencies with different decay times were observed.}
 \label{fig:seo}
\end{figure}

In order to investigate the nuclear spin-spin coupling in NbP, we have measured the $^{93}$Nb-NMR spin-echo decay profile at several temperatures for $\mu_{0} H \parallel [0 0 1]$. The obtained spin-echo decay and oscillation part at 4.3\,K are shown in Fig.\,\ref{fig:seo}(a) and (b). The spin-echo decay profile obtained by changing the time between the first exciting and the second refocusing (delay time). It exhibits a simple exponential decay with noticeable oscillation. The oscillation part is shown in Fig.\,\ref{fig:seo}(b) where we see two different oscillations. A long-period oscillation with $\tau_{\rm p,long} = 20.4$\,$\mu$s ($f_{\rm p,long} = 49.0$\,kHz) appears in addition to the short period oscillation with $\tau_{\rm p,short} = 3.08$\,$\mu$s ($f_{\rm p,short} = 325$\,kHz). $f_{\rm p}$ is related to the coupling strength between the nuclear spins. Such oscillation is often observed in NMR with quadrupolar perturbations \cite{Abe1966}. In such a case, the oscillation frequency is determined by the nuclear quadrupolar frequency for given field direction. In the present case, the quadrupolar frequency for $\mu_{0} H \parallel [0 0 1]$ is calculated to be $\nu_{{\rm Q}, 0 0 1} = 250$\,kHz by using $\nu_{\rm Q} = 0.61$ MHz and $\eta = 0.20$ and EFG directions as determined above. Thus, the fast-oscillating component is well explained by the nuclear quadrupolar coupling.

On the other hand, the slowly oscillating component should have a different origin. A similar oscillation of the spin-echo intensity is observed by $^{181}$Ta-NQR in TaP and the relations to the Weyl fermion excitations are discussed \cite{Kubo2023}. However, $f_{\rm p}$ in NbP is about 10 times larger than the spin-echo oscillation in TaP ($f_{\rm p} = 3.58$\,kHz). The indirect nuclear spin-spin coupling via virtual excitation of the Weyl fermions may be a plausible origin of the oscillation, but the real origin is not definitively determined at this time. Future measurements of anisotropy and magnetic field dependence would be interesting.

\section{Concluding Remarks}
Expanding our previous $^{181}$Ta-NQR works on TaP and TaAs \cite{Yasuoka2017,Kubo2023}, in this article we have presented an extended microscopic study in one of the prototype Weyl semimetals, NbP, using the $^{93}$Nb-NMR technique. From the full angular dependence of 18 quadrupole split NMR lines in a single crystal, we have extracted the anisotropic EFG and Knight shift tensors. The NQR parameters agree well with the DFT calculated EFG tensors. The Knight shift tensor was also obtained. We also studied the Knight shift as a function of temperature $K (T)$ and external field $K (\mu_{0} H)$. $K (T)$ decreased linearly with decreasing temperature like the bulk susceptibility, while we found that $K (\mu_{0} H)$ has a clear oscillation at low temperatures. This oscillation is associated with the quantum oscillation due to nearly massless Weyl fermion bands, and the frequency is consistent with that obtained from bulk measurements. This is one of the first example to show quantum oscillations in the Weyl fermion system by NMR.

We also report on the temperature dependence of the nuclear spin-lattice relaxation rate ($1 / T_{1} T$) and the spin-echo decay profile at 4.3\,K. These results have a contrast between TaP and TaAs \cite{Yasuoka2017,Kubo2023}. In TaP and TaAs, both the relaxation behaviors are well documented by the Weyl fermion excitations with a temperature-dependent orbital hyperfine interaction for the TaP case. In NbP, however, we have an essentially temperature independent process and qualitative agreement with the non-interacting Fermi liquid scenario suggested by the DFT band structure.

Two-component spin-echo oscillations have been observed. The shorter-period oscillation is identified to be due to the origin of quadrupole coupling, while the longer-period oscillation may indicate the presence of indirect nuclear spin-spin coupling, as discussed in TaP. Notably, the oscillation period in NbP is much shorter than that observed in TaP, and further studies are needed to understand this difference.

Considering the partial uncertainties in our current interpretations, we remain optimistic that conducting further investigations using NMR/NQR techniques could provide further insights into the physical properties of topological Weyl semimetals. In particular, detailed studies of the field dependence and anisotropy of the Knight shift, $1 / T_{1}$, and $1 / T_{2}$ shed light on the nature of the Weyl fermions and their interactions with the nuclear spins.

\section*{Acknowledgements}
We would like to thank Vicky Hasse for support
during the synthesis, Steffen H\"{u}ckmann for
collecting the powder X-ray diffraction data as well
as Dr. Ulrich Burkhardt and Petra Scheppan for
EDX investigations.

\section*{Appendix. Details of the NMR spectral analysis}
To understand the complete angular dependence of the NMR spectra, we follow the procedure described in Ref. \cite{Kubo2020}.

Here, we introduce the following four coordinates to analyze the complex angular dependence of $^{93}$Nb-NMR spectra in NbP: (1) laboratory coordinate $\Sigma_{\rm L}$, in which $\mu_{0} H$ is along with the $z$-direction, (2) stage coordinate $\Sigma_{\rm S}$ with its $y$-axis is the rotation axis and its $z$-direction is perpendicular to the stage plane, (3) crystal coordinate $\Sigma_{\rm C}$ where $[1 0 0]$, $[0 1 0]$, and $[0 0 1]$ directions of the crystal are $x$-, $y$-, and $z$-axis, respectively, and (4) principal coordinates $\Sigma_{\rm P}$ in which $\hat{V}$ and $\hat{K}$ are diagonal. Next, we define the
basis transformation from $\Sigma_{i}$ to $\Sigma_{\rm j}$ by the rotation matrix $R_{i j} (\alpha_{i j}, \beta_{i j}, \gamma_{i j}) \equiv R_{z} (\alpha_{i j}) R_{x'} (\beta_{i j}) R_{y''} (\gamma_{i j})$. $R_{z} (\alpha_{i j})$ is the $z$-rotation by the angle $\alpha_{i j}$, etc. The $x'$
-axis refers the direction after $z$-rotation, etc. $(\alpha_{i j}, \beta_{i j}, \gamma_{i j})$ are the Tait-Bryan angles that relate to two coordinates.

The relationship between these angular parameters and experimental conditions are:

1. The external field angle $\theta$ is described by $\theta_{\rm LS}$ as $R_{\rm LS} (0, 0, \theta)$.

2. The imperfection of the sample mounting is
described by $R_{\rm SC} (\alpha_{\rm SC}, \beta_{\rm SC}, \gamma_{\rm SC})$. For the perfect mounting as shown in Fig.\,1, the matrix is $R_{\rm SC}(0, 0, 90^{\circ})$.

3. The principal axis directions can be estimated by the local symmetry of the Nb site, the anisotropy of the spectrum, and the band calculation. That is given by $R_{\rm CP} (0, 0, \pm 90^{\circ})$ for the Nb1 site and $R_{\rm CP} (\pm 90^{\circ}, 0, \pm 90^{\circ})$ for the Nb2 site.

By using these matrices, $\hat{V}_{\rm L}$ and $\hat{K}_{\rm L}$, in $\Sigma_{\rm L}$, are given by
\begin{align}
 \hat{V}_{\rm L}
 &=
 R_{\rm LS} R_{\rm SC} R_{\rm CP} \hat{V}_{\rm P} R_{\rm CP}^{- 1} R_{\rm SC}^{- 1} R_{\rm LS}^{- 1} \\
 \hat{K}_{\rm L}
 &=
 R_{\rm LS} R_{\rm SC} R_{\rm CP} \hat{K}_{\rm P} R_{\rm CP}^{- 1} R_{\rm SC}^{- 1} R_{\rm LS}^{- 1}
\end{align}
where $\hat{V}_{\rm P}$ and $\hat{K}_{\rm P}$ are diagonal in $\Sigma_{\rm P}$. The parameters given in the main text and solid lines in Fig.\,\ref{fig:peaks} are the least-squares fitting to the experimental peak fields.

\end{document}